\documentclass[reprint,
nofootinbib,
 amsmath,amssymb,
 aps,
 prd,
]{revtex4-2}

\usepackage{graphicx}
\usepackage{dcolumn}
\usepackage{bm}

\usepackage{color,ulem}
\usepackage{natbib}

\newcommand{\mpl}{m_{\rm pl}}
\newcommand{\epl}{E_{\rm pl}}
\newcommand{\kb}{}

\begin{document}

\title{Black Holes as ``Time Capsules": A Cosmological Graviton Background and the Hubble Tension}

\author{Tsvi Piran}
 \email{tsvi.piran@mail.huji.ac.il}
 \affiliation{Racah Institute for Physics, The Hebrew University, Jerusalem 91904, Israel}
\author{Raul Jimenez}
 \email{raul.jimenez@icc.ub.edu}
\affiliation{ICC, University of Barcelona, Marti i Franques 1, 08028 Barcelona, Spain.}
\affiliation{ICREA, Pg. Lluis Companys 23, 0810 Barcelona, Spain.} 

\date{\today}

\begin{abstract}
Minuscule primordial black holes before the end and after inflation can serve as ``time capsules" bringing back energy from the past to a later epoch when they evaporate. As these black holes behave like matter, while the rest of the Universe content behaves like radiation, the mass fraction of these black holes, that is tiny at formation, becomes significant later. If sufficiently small, these black holes will evaporate while the Universe is still radiation dominated. We revisit this process and point out that gravitons produced during the evaporation behave as  ``dark radiation". If the initial black holes are uniformly distributed so will be the gravitons  and in this case they will be  free of Silk damping and avoid current limits on ``dark radiation" scenarios. Seeds for such black holes can arise during the last phases of inflation. We show here that with  suitable parameters, this background graviton field can resolve the Hubble tension. We present current observational  constraints on this scenario and suggest upcoming observational tests to prove or refute it. Finally, we also elaborate on the graviton background produced by particle annihilation during the Planck era or shortly after inflation. 
\end{abstract}
\maketitle

\section{Introduction}
\label{sec:intro}

Hawking's ~\cite{Hawking1974,Hawking} dramatic discovery of black hole (BH) evaporation was the first link between general relativity and quantum theory, the two pillars of physics. In one of these papers \cite{Hawking1974}  Hawking predicts that primordial BHs of $10^{15}$g will appear as cosmic explosions today. However, in spite of the almost 50 years that have passed, BH evaporation has evaded any observational confirmation\footnote{See, however, claims of observations of Hawking radiation in analogue systems \cite{Weinfurtner2011,analogue}}. We explore here potential cosmological consequences of BH evaporation  and show that it might have had dramatic effects on the early Universe. 

Minuscule BHs that form in the early Universe evaporate eons later. The surrounding radiation energy density decreases like  $(1+z)^{4}$ while the BHs' energy density decreases like $(1+z)^{3}$. Thus, a tiny fraction of the early Universe mass captured in such BHs can be significant or even  dominant by the time they evaporate. As such, these BHs can be considered as ``time capsules" that carry energy\footnote{If BH evaporation releases the information captured within the BH at formation, in principle, these BHs can served as ``time information capsules" as far as information as well. However, capturing the exact phases of all the radiated particles, which is essential for that, might prove very challenging as it will require a cosmic Bell-like experiment.}  from their formation time and deposit it much later.

The original idea was proposed by Hopper et al., ~\cite{Hooper}  and various
aspects of this process have already been explored in detail (see e.g. \cite{Masina,Arbey+} and references therein and the review in~\cite{Auffinger}). Here we propose an extension of this scenario in which we explore the implication of a possible {\em homogeneous} cosmological graviton background (CGB) as to provide the means to address the Hubble tension. 

This involves the possibility that  minuscule BH ($m_{_{\rm BH}} <  10^{12} $g and typically much smaller) that form during  inflation provide a source of pure ``dark radiation" in the form of gravitons. This dark radiation can  change the Friedman equation at recombination and thus, it can  solve the Hubble tension (see e.g. \cite{VerdeTreu} and references therein). 

It is worth mentioning that the current constraints on $\Delta N_{\rm eff}$ from the Planck satellite combined with Big Bang Nucleosynthesis (BBN) and large scale structure~\cite{Planck18} limit it to be $< 0.25$ while a value of $\approx 0.4$ is needed to fully remove the tension. However, the above constraint comes from the Silk damping and perturbation effects on the CMB high multipoles and  does not apply to the Friedman equation as we will show in \S~\ref{sec:tension}. Thus any background that has no perturbation and Silk damping will avoid these constraints~\cite{NilsBBN,BBNNils}.

 We consider first, in \S~\ref{sec:formation},  the formation  and the energy budget of these  minuscule BHs. In \S~\ref{sec:evaporation} we discuss the composition of particles emitted and their thermalization and  the homogeneity of this process. We address a possible solution to the Hubble tension puzzle in \S~\ref{sec:tension}. We conclude in \S~\ref{sec:conclusions} with a brief summary and a discussion of possible observational signatures.  As the formation mechanism (if any) of BHs at the early Universe is highly uncertain, our estimates are given only up to  factors of order unity. In an appendix we discuss aspects of graviton formation via annihilation processes during or shortly after the Planck era, and the corresponding CGB production. While $\Omega_{\rm g}$ produce in this way is probably too small to resolve the Hubble tension, this is an interesting route to produce CGB whose nature could be used to explore  the quantum nature of gravity.
\section{Formation}
\label{sec:formation}

Consider a BH that evaporates at redshift ${z_{\rm ev}}$. 
The evaporation takes place over a time scale $H^{-1}(z_{\rm ev})$, the inverse of the Hubble parameter at $z_{\rm ev}$. Equating  $H^{-1}$  to the evaporation life-time of the BH, $5120 \pi G m_{_{\rm BH}}^3 /(c^3 \mpl^2)$,  we find $m_{_{\rm BH}}(z_{\rm ev})$,  the mass of a BH that evaporates at ${z_{\rm ev}}$: 
\begin{equation} 
m_{_{\rm BH}} (z_{\rm ev})  \approx 0.02~ \mpl~ \big( \frac{  \epl  }{\kb T}\big)^{2/3}   \approx   5\cdot  10^{8} ~{\rm g}~ \big(\frac{{\rm MeV} }{T}\big)^{2/3}  \ , \label{eq:mbh}
\end{equation}
where $\mpl$ and $\epl$ are the Planck mass and Planck energy respectively and  $T$ is the  temperature of the Universe  at ${z_{\rm ev}}$.

\begin{figure}
\vskip.3cm
\includegraphics[scale=0.38]{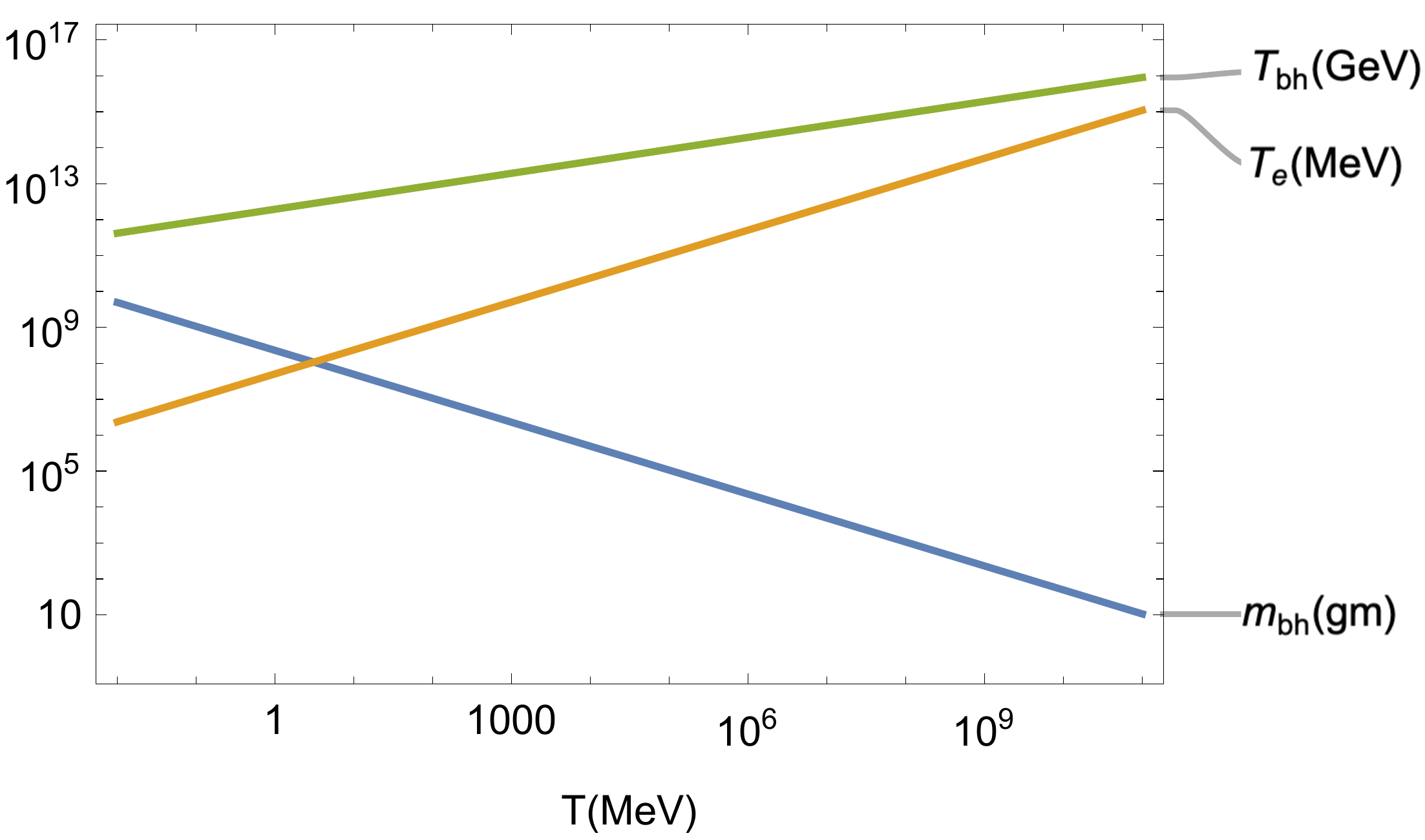}
\caption{$m_{_{\rm BH}}$, Mass of the BHs (in grams - blue line), $T_e$ the  temperature  of the evaporating BHs (in MeV - orange line)  and $T_{_{\rm BH}}$  the background temperature of the Universe (in GeV - green line) when these  BHs form as a function of $T$, the Universe temperature at the time the BHs evaporate.}
\label{fig:TM}
\end{figure}

\begin{figure}
\vskip.3cm
\includegraphics[scale=0.44]{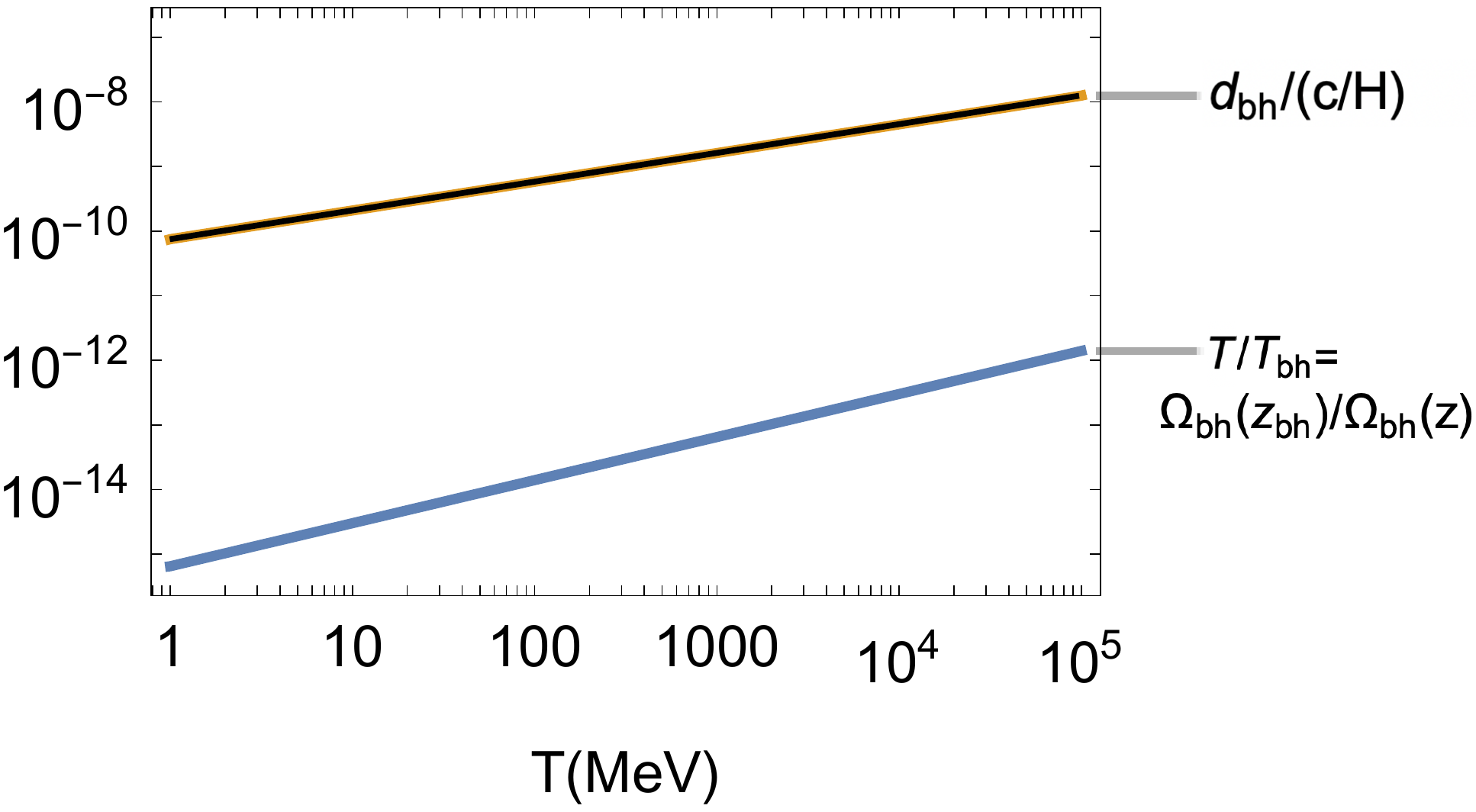}
\caption{The ratio, $d_{_{\rm BH}}/(c/H)$, of the distance between the evaporation BHs to the horizon (black)  and the ratio,$T/T_{_{\rm BH}}$,   of the  temperature at BH evaporation, T,  to the  temperature at formation, $T_{_{\rm BH}}$ (blue),   as a function of $T$. The latter ratio also equals (see Eq.  \ref{eq:zbh}) the ratio $\Omega_{_{\rm BH}}(z_{_{\rm BH}})/\Omega_{_{\rm BH}}(z_{\rm ev})$, demonstrating that a minute fraction of the energy density of the Universe in BHs at the early Universe can be very significant and even dominant later. }
\label{fig:Ratios}
\end{figure}

The horizon mass of a flat Universe is comparable to a BH mass of the same size. This determines the formation epoch $z_{_{\rm BH}}$ of the BHs  evaporating at $z_{\rm ev}$ (see Fig.~\ref{fig:TM}). In turn, this determines the temperature ratios and the ratio between the fraction of the BH energy densities at evaporation, $\Omega_{_{\rm BH}}(z_{\rm ev})$, and at formation, $\Omega_{_{\rm BH}}(z_{_{\rm BH}})$   (see Fig.~\ref{fig:Ratios}):
\begin{equation} 
\frac{z_{_{\rm BH}} (z_{\rm ev}) }{z_{\rm ev}} = \frac{\Omega_{_{\rm BH}}(z_{\rm ev})}{\Omega_{_{\rm BH}}(z_{_{\rm BH}})}  \approx 3~\big( \frac{   \epl }{\kb  T} \big)^{2/3} \approx 1.5 \cdot 10^{15} \big(\frac{{\rm MeV} }{T}\big)^{2/3} \ . 
\label{eq:zbh}
\end{equation} 
The corresponding  temperature of the Universe at the time that the BHs form , $T_{_{\rm BH}}$, is (see Fig.~\ref{fig:TM}):
\begin{equation}
T_{_{\rm BH}} (z_{\rm ev})   \approx 3~ \epl \big( \frac{\kb  T} {   \epl } \big)^{1/3}
 \approx  3 \times 10^{13} ~{\rm GeV} \big(\frac{T}{{\rm MeV} }\big)^{1/3} \ .
\label{eq:Tbh}
\end{equation}

Typical values are 
$z_{_{\rm BH}}=10^{23}$ and $m_{_{\rm BH}}= 10^8$g for evaporation just before nucleosynthesis.   The temperature of the Universe, $T_{_{\rm BH}}$, at the time that these BHs form is  $\approx 10^{12}$ GeV,  just a few orders of magnitude  below the canonical scale of exit from inflation. This suggests that  their formation  may be  associated with the end of inflation. Specifically, it is possible that exit from inflation isn't smooth and the fluctuations that led to the formation of minuscule  BHs arose at this stage, or that BH formation is a generic feature at small scales~\cite{Fisher}. 

After formation, BHs will accrete some of the surrounding matter.  With $\dot M = \pi G^2 m_{_{\rm BH}}^2 \rho/c_{\rm s}^3$, where $\rho, c_{\rm s}=c/\sqrt{3}$ are the surrounding density and speed of sound,
this accretion is significant only right after BH formation when the surrounding density is largest. At this stage the BHs grow by a factor of $\sim 2$ compared to their initial mass. This order of unity effect can be  ignored. 

\section{\label{sec:evaporation} Evaporation}

\subsection{The Composition of the evaporating particles. }

The  evaporation temperature, $T_{\rm e}$:  
\begin{equation}
T_{\rm e} \approx 2~ \big( \frac{ \mpl}{\kb T }\big)^{1/3} ~ T 
\approx 10^{7} {\rm GeV}~ \big(\frac{T}{{\rm MeV} }\big)^{2/3} \ , 
\label{eq:tinj}
\end{equation}
determines the composition of the particles  that are produced according to their rest mass and spin \cite{Page1976}. {\it Photons, neutrinos and gravitons}, that are massless\footnote{Neutrinos are effectively massless at the relevant temperatures discussed}, are produced at any temperature. Their energy density ratio depends strongly on the evaporating BH spin. Gravitons dominate  for high spin BHs and neutrinos for low spin ones  \cite{Page1976}.
In the following we assume, lacking information on these primordial  BHs, that their spin distribution results in roughly equal amounts of  different massless particles. 

At higher evaporation temperatures massive particles form. Within most of the relevant regime,  evaporation temperatures are   above the QED and QCD phase transitions and the composition of the evaporating matter resembles the early Universe composition with one significant difference, production of thermal gravitons. At these temperatures $g_*$
that measures the effective number of degrees of freedom\footnote{Considering only the degrees of freedom of the standard model of particle physics at those energies.} (see e.g. Fig. ~1 in \citep{Borsanyi2016}), is $\approx 100$,
and at first glance is seems that   $f_{g}$, the fraction of the BH energy deposited in gravitons, is   small  ($f_g \sim 1$\%) \cite{Page1976}. However this fraction increases if the BHs are  rotating rapidly \cite{Page1976R}.
$f_g$ can be as high as  $10$\% for $a > 0.95$ and seems to plateau at that value even for $a \sim 0.9999$ 
\cite{Arbey+}. 

The density in gravitons with respect to the whole radiation one is~\footnote{At high temperatures (corresponding to larger than matter-radiation equality) the entropy degrees of freedom are equal to the particle ones.}
\begin{equation}
    \frac{\rho_G (T_{\rm eq})}{\rho_R (T_{\rm eq})} = f_g \left ( \frac{g_s (T_{\rm eq} )}{g_s (T_{\rm ev})} \right )^{1/3} 
\end{equation}

\subsection{Thermalization}
\label{sec:thermalization}

The fate of the produced particles depends on their interaction with the surrounding matter. Within the relevant temperature range that we consider here ($T \gtrsim 1$ MeV) all electromagnetically or strongly and even weakly interacting particles are strongly coupled and they thermalize quickly, depositing their energy to the rest of the energy reservoir  at that time. However, Gravitons remain decoupled at all temperatures. As gravitons propagate freely  
they form  a background radiation field whose typical energy today is   $T_{\rm e}/z_{\rm ev}$ (see Fig. \ref{fig:TM}): 
\begin{equation}
h \nu_{g} \approx \big( \frac{ \mpl}{ T }\big)^{1/3} ~ T_0 
\approx 10  ~ {\rm keV } ~\big(\frac{{\rm MeV}}{T} \big)^{1/3} \ , 
\label{eq:today}
\end{equation}
where $T_0$ is the current CMB temperature. 
These energies corresponds to  frequencies of $\approx  10^{18}$Hz. These are  comparable to the energies of gravitons produced by gravitational  Bremstrahlung at the Sun \cite{Weinberg1965}. This prediction can, at least in principle, be tested observationally. 

{\it Massive relics:} Over most of the relevant range of $T$ the evaporation energy, $T_{\rm e}$,  is much larger than the QCD and QED transition energies.    The mean free path of the vast majority of these particles will be much smaller than the horizon size and they quickly thermalize. An interesting exception is a case in which a particle, denoted $X$,  has the lowest mass with a specific charge. At  $T< m_{X}$ particle $X$ can only annihilate with its anti-particle, $\bar X$. Depending on the details of the evaporation and on $\sigma_{X\bar X}$  $\Omega_X$ might be significant. Importantly, $\Omega_X$ won't satisfy, in such a case the common relation (the so-called WIMP miracle) between $m_X$ and $\sigma_{X\bar X}$ that hold for a regular freeze-out. This leaves room for the formation of a WIMP type of dark matter that doesn't satisfy this condition, a possibility that has been  explored further elsewhere~\cite{Hooper,Masina,Arbey+,Profumo}. 

\subsection{Black Hole Scale Homogeneity  }
\label{sec:homogeneity}
The  evaporating BHs  masses are extremely small compared to $M_{\rm H}(z_{\rm ev})$, the  horizon  mass at evaporation: 
\begin{equation}
\frac{m_{_{\rm BH}}(z_{\rm ev})}{M_{\rm H}(z_{\rm ev})}\approx  0.01~ \big(\frac{ \kb T} {\epl}\big)^{4/3}\ \approx 4 \cdot 10^{-31} \big(\frac{  T} {{\rm MeV}  }\big)^{4/3}.
\end{equation} 
The corresponding ratio between the distance between evaporating BHs, $d_{_{\rm BH}}$ and the horizon $c/H$,   determines the inhomogeniety (or lack off) induced by the evaporation:
\begin{equation}
\frac{d_{_{\rm BH}}}{c/H(z_{\rm ev})} \approx 0.1 \big(\frac{\kb T}{ \epl}\big)^{4/9} 
\approx   10^{-9}
\big(\frac{T}{{\rm MeV} }\big)^{4/9}  \ .
\label{eq:evaporation}
\end{equation}
As this ratio is extremely small (see Fig.~\ref{fig:Ratios})
the   inhomogeneity that it introduces is erased on a time scale much shorter than the horizon crossing time
and  the evaporation does not affect  the homogeneity of the Universe. 

\section{The Hubble Tension}
\label{sec:tension}

The recently realized ``Hubble tension" between early and late Universe observations may indicate or point to the need of new physics to describe the Universe (see e.g. the review in~\cite{VerdeTreu} and references therein). 
This tension consists of a mismatch between the model-dependent inferred value of $H_0$ from the cosmic microwave background (CMB) temperature and polarization data~\cite{Planck18} and direct measurements using parallaxes and different standarizable candles in the local Universe ($z<0.1$), which are cosmology independent. The disagreement is at the $4-5 \sigma$ level~\cite{VerdeTreu} and translates into the local value of $H_0$ being about 10\% larger than the CMB inferred one. It is worth taking it seriously given the exhaustive tests for systematic uncertainties that both measurements have endured, finding no obvious source for such an effect.

 There is already a myriad of theoretical solutions proposed (see~\cite{olympics} for a fairly exhaustive description of possible solutions) to alleviate and resolve this tension. All boil down to the need to change the anchors (either at the early Universe or the late one) as the evolution since the high-z Universe ($H(z)$ for $z < 2$ as given by SN, BAO and cosmic chronometers) to the local one is well described by the current LCDM paradigm~\cite{ruler,bernal}. For the early Universe this translates into changing the length of the standard ruler. 

It is straightforward to see that a change in the energy densities will lead to a change in the inferred $H_{0}$ value from the CMB. To establish the value of $H_{0}$ we use the CMB to infer angles and the sound horizon scale, $r_s$: 
\begin{equation}
\label{eq:rs}
    r_{s} = \int_{z_{s}}^{\infty} \frac{c_{s} dz}{H(z)} \ .
\end{equation}
This equation is fully model dependent as it relies on the assumed value for the sound speed $c_{s}$, the recombination redshift $z_{s}$ and the Hubble parameter $H(z)$
A change in the Hubble parameter up to the recombination redshift will change the value of $r_s$. In particular, an increase of $H(z)$ by some additional radiation field will decrease $r_s$, so the true value of $H_{0}$ would have to be higher in order to compensate for this change of length. This solution is generally refered to as ``dark radiation". 

Usually ``dark radiation" is discarded as a solution to the Hubble tension because in these models Silk damping of this dark radiation field puts a strong limit  the amount of dark radiation due to CMB constraints~\cite{Planck18}. 
This limits is expressed using $N_{\rm eff}$, the extra relativistic degrees of freedom affect the thermal budget of the Universe according to the well known equation (see e.g.~\cite{KolbTurner})
\begin{equation}
\Delta N_{\rm eff} = \frac{4}{7} g_{Y,*} \left ( \frac{43}{4 g_{*} (T_{F})} \right )^{4/3}  
\end{equation}
where $g_{*}$ is the effective number of degrees of freedom of the dark radiation particle $Y$ and $g_{*} (T_{F})$ is the effective number of relativistic degrees of freedom in {\it thermal equilibrium} at the
temperature $T_{F}$ at which $Y$ decouples from the plasma.
To resolve the tension we need to increase 
$H_0$ to change the sound horizon scale as can be trivially seen in~(\ref{eq:rs}).  Within ``dark radiation" solutions this corresponds to increasing $N_{\rm eff}$ by  $\approx 0.4$.

The problem is that the value of $\Delta N_{\rm eff}$ needed to fix the tension ($ \approx 0.4$~\cite{VerdeTreu}) is already ruled out by observations.    Silk damping limit $\Delta N_{\rm eff} < 0.25$ (e.g.~\cite{olympics}).  
A second, weaker limit on $\Delta N_{\rm eff} $  arises from  BBN constraints, as the combination of deuterium and helium, limits the value of $N_{\rm eff}$ in a nearly cosmology model independent way.
Using their Fig.~8 of Ref.~\cite{BBNNils}, the right panel shows that indeed, it is possible to have { $\Delta N_{\rm eff} \lesssim 0.5$.} With this upper limit we obtain values of $H_0$ as high as 73 km s$^{-1}$ Mpc$^{-1}$ within the current constraints of BBN at the 95\% confidence level. This will remove the tension. 

Previous estimates of the possible role of massless particles produced by evaporation primordial BHs \cite{Hooper} focused on  Silk damping  and concluded that, since such particles can contribute at most $\Delta N_{\rm eff} \approx 0.2$ they cannot resolve the tension. However, this limit can be avoided if the ``dark radiation" does not suffer from Silk damping. This can happen for gravitons produced by evaporating miniscule BHs if the BHs are distributed uniformly. As the gravitions don't interact with the rest of the matter they will remain uniform. In this case the only limit on ``dark radiation" will be the BBN limit which allows a solution of the Hubble tension.

The question is whether enough gravitons are produced so that their  energy density, $\Omega_{\rm gr}$, 
(corresponding to $\Delta N_{\rm eff} \sim 0.4$) is 10\% of the CMB energy.
Using eq. 32 in~\cite{HooperKerr}: 
\begin{equation} 
\Delta N_{\rm eff,g} \sim 0.013 \big(\frac{f_{g}}{0.0047}\big) \big(\frac{106}{g_{*}}\big)^{0.33} \Omega_{\rm BH}(z_{\rm ev})  \ , 
\end{equation} 
If evaporation is before BBN\footnote{The high energy photons arising from evaporation after BBN may destroy the BBN products.} then $g \sim 100$. If the BHs are fast rotating then $f_{g} = 0.1$ and $\Delta N_{\rm eff,G} \Omega_{\rm BH}(z_{\rm ev}) \sim 0.3 \Omega_{\rm BH}(z_{\rm ev})$, which is consistent, although an underestimate, with the values found in the detailed numerical analysis by~\cite{Masina,Arbey+}. 
Thus, to produce sufficient gravitions the BHs must be rapidly rotating and $ \Omega_{\rm BH}(z_{\rm ev})$ must be of order unity.

The above considerations puts somewhat opposing  limits on the conditions needed to resolve the Hubble tension. On one hand  $\Omega_{\rm BH}$ must be of order unity. On the other hand, these uniformly distributed   BHs produce other particles as well. Hence, they shouldn't be too abundant as otherwise their products would  dominate the Universe resulting in a uniform Universe. These two opposing conditions suggest a small window of $\Omega_{\rm BH} \sim 0.8-0.9$
at the time of evaporation. Within this enough gravitons are produced. The other products of the evaporating BHs  dominate, but there is enough room for  relic material to  produce the needed structure. Because of the dilution factor due to the evaporation product this relic must to have higher fluctuations than what is observed today.  

Let us summarize the conditions needed for gravitons produced by evaporating BHs to provide a solution to the Hubble tension:
\begin{enumerate}
\item BHs must form uniformly\footnote{With these conditions dark matter cannot be formed from the evaporating BHs as it must be non-uniform.}. 
Within the context of inflationary models this could happen if the BHs for during the last epochs of inflation or right after it during the re-heating phase. Recall that only a minute fraction of the Universe is initially in these BHs. 
\item BHs must be rotating with spins $>0.9$ as to provide a sizable amount of gravitons. 
\item $\Omega_{\rm BH}$ must be close to unity but somewhat smaller as to avoid all energy content in the universe to be homogeneous. This means that most of the matter in the Universe, but not all, has been ``recycled" through black holes. The condition $\Omega_{\rm BH}(z_{\rm ev}) \approx 0.8-0.9 $ can be seen as a fine-tuning requirement in our model.
\item Evaporation must take place before BBN as to not alter the succesful agreement of current BBN estimates with observations. This  constrains the maximal mass of the BHs to $< 10^{9}$g.
\end{enumerate}

\section{Conclusions and Observational Signatures}
\label{sec:conclusions}
Light ($M_{\rm BH} < 10^{12}$g) primordial BHs will evaporate during the radiation dominated era and those with $M_{\rm BH} < 10^{9}$g before BBN. Most of the evaporating particles, apart from  gravitons and possibly some unknown  weakly interacting particles  that may contribute to the dark matter, thermalize rapidly. 
A background field of $\sim10^{18} $Hz gravitons is  the natural remnant of these BHs. With suitable parameters the resulting  graviton radiation field can resolve the so-called Hubble tension. 
This can happen if the BHs are (i) rapidly rotating,  (ii) uniformly distributed and (iii) at the time of evaporation they compose a dominant fraction, $\sim 0.8-0.9$, of the energy density of the Universe .  

In our scenario seeds for primordial black holes are formed  during inflation and  inflation homogenizes their distribution. Our main assumption is that these BHs  rotate rapidly as in order to produce gravitons. Although even almost non-rotating BHs do produce some gravitons, so in this case their abundance will need to higher by a factor $\sim 10-100$. One important point is that the required initial energy density in  these BHs is so tiny $<10^{-10}$ or less, that even if their seeds are produced during the late phases of inflation and hence are diluted by expansion they will have a noticeable effect.

The  observational predictions of these scenarios are very clear: First, gravitational wave experiments in the $10^{18} $Hz range  should see a stochastic background. If this is indeed the solution of the Hubble tension, then the energy density of this background is a significant fraction (a few percent) of the CMB energy density. Exploring this window (see e.g.~\cite{Blas} and references therein) could therefore also unveil the presence in the early Universe of evaporating BHs and indirectly probe the quantum nature of gravity.

While detecting these gravitons is an experimental challenge for the future, the above estimates provide an avenue to test our scenario rather soon. Improvements in theoretical reactions rates relevant to BBN calculations by the LUNA~\footnote{\text{https://luna.lngs.infn.it}} project, as well as further analysis from large scale structure by the ShapeFit team~\cite{shapefit}, will provide a much tighter constrain on the allowed value of $\Delta N_{\rm eff}$ during BBN. This can either refute   our scenario or support it. Strong upper limits on $\Delta N_{\rm eff}(BBN) $ would put very tight constraints on primordial BH production with $M_{\rm BH} \lesssim 10^9 $g/cm$^3$.

\begin{acknowledgments}
 Funding for this work  was partially provided by (TP) an Advanced ERC grant TReX and   by (RJ) project PGC2018-098866-B-I00, FEDER “Una manera de hacer Europa” and “Center of Excellence Maria de Maeztu 2020-2023” award to the ICCUB (CEX2019-000918-M funded by MCIN/AEI/10.13039/501100011033). One of us (RJ) thanks coffee house ``fin del mondo" in Piran (Slovenia) where some zoom discussions concerning this work took place.   
\end{acknowledgments}

\vspace*{0.5cm}

\appendix
\section{Cosmological Graviton Production via Annihilation}

\label{sec:prod}

One of the firm predictions of quantum gravity is that quanta known as gravitons should exist. Gravitons are quantum whereas the detected gravitational waves by LIGO are classical. To verify that gravitons exist
we must detect single gravitons, for example by observing
gravito-ioninization atomic transitions due to absorption of a graviton - or an equivalent type of quantum experiment.

In the main body of this paper we have explored the graviton production via the super-radiance process in Hawking evaporation. There is, however, a much less explored process that can lead to a measurable signature of the quantum nature of gravity. This process is the production of gravitons via annihilation. The annihilation  process is  quantized  producing  two gravitions in each event.  Even an  indirect evidence for the existence of this particular CGB would verify that gravitons  do exist as quantum objects and we consider the detection prospects.

In this appendix we explore a different route to produce a uniform cosmological gravitation background. This is by the process 
\begin{equation}
X \bar X \rightarrow 2~g\ ,
\end{equation}
where $X$ is an arbitrary particle. 
The annihilation  cross section  to gravitons (denoted hereafter $g$) is extremely small but non-vanishing. Detailed calculations (\cite{Vladimirov1963}, see \cite{Papini1977} for a review),  as well as dimensional analysis, suggest that it behaves like, 
\begin{equation}
\sigma_{x\bar x \leftrightarrow g g } \approx  (G E /c^4)^2=r_g^2 (E) \ , 
\end{equation} 
the square of the gravitational radius of a particle with a total energy $E$. Thus,  two particles with energy $E$ can annihilate and produce two gravitons of the same energies. As the cross section is proportional to the square of the gravitational radius of the particles, this process is significant only when the typical energy of the particles is of order of the Planck energy. Similarly the inverse process of annihilation of two gravitons to a particle-antiparticle pair will be significant only at the Planck era.  Namely it is only at this energy that gravitons can be in thermal equilibrium with other particles. 

Once formed, the graviton's energy decreases like the expansion factor $a^{-1}$. As the energy of the gravitons decreases their cross section for annihilation decreases rapidly. Thus, apart from a brief phase at the Planck era gravitons can never be  in thermal equilibrium with other matter fields \cite{Zel'dovich1967,Smolin1985}. For this reason, Zel'dovich \cite{Zel'dovich1967} 
remarked that the  graviton background is
determined just by the  initial conditions of the Universe.  
This is in contrast with thermal equilibrium at lower temperatures,  assumed  by ~\cite{Matzner} and subsequent work, that have led to an overestimated $\Omega_{\rm g}$.
Once gravitons form their chance to annihilate back to other form of matter is negligible. The CGB is an energy sink that hides a fraction of the total energy as dark-radiation. Namely, it has no interaction apart from gravity and  will only contribute to the radiation energy density of the Universe.

The graviton energy production rate per volume element in a thermal bath with a temperature $T$ and a density $n\approx (k T /\hbar c)^3$
is: 
\begin {equation} 
\dot e_{\rm gr} \approx c n^2 r_g^2(E)  E = \frac{G^2}{c^7} n^2 E^3 \ ,
\label{eq:er}
\end{equation}
where $E = kT$.
In a Hubble time, $ H^{-1}  \sim ( G e/c^2)^{-1/2}$, an energy $\dot e_{\rm gr}H^{-1}$ is converted per comoving volume to gravitons. Its ratio to the total energy density is: \begin{equation}
\Omega_{\rm g} = \frac{\dot e_{\rm gr} H^{-1}}{ e  } \approx \bigg( \frac{E}{E_{\rm pl}}\bigg)^3 =\bigg( \frac{T}{T_{\rm pl}}\bigg)^3 \ ,
\label{Egene}
\end{equation}
where $E_{\rm pl}$ is the Planck energy. This last relation can be cast into the form \begin{equation}
\Omega_{\rm g} \approx \bigg(\frac{E_{\rm gr}}{E_{\rm pl}}\bigg)^3~ \Omega_{\rm CMB} \ ,
\label{Omega}
\end{equation} 
where $E_{\rm gr}$ is the energy scale when most gravitons formed and $\Omega_{\rm CGB;CMB}$ are the cosmological energy fractions. This ratio is of order unity for the Planck era but it decreases rapidly as the temperature decreases.

We  have to consider two different scenarii distinguished by the question of whether inflation took place or not. 

\subsubsection{Annihilation CGB with Inflation}
Inflation that took place  sometime after the Planck era, diluted and erased the CGB that formed during the Planck era. As inflation ends a thermal phase is restored at a temperature $T_{\rm inf}$. 
A minuscule fraction of the particle will annihilate to gravitons, forming a new CGB, whose fraction of the total energy is $\approx (T_{\rm inf}/T_{\rm pl})^3$. The initial energy of these gravitons will be $T_{\rm inf}$, but of course they won't be in thermal equilibrium with the rest of the universe and their density will be much lower than the thermal density of massless particles at this temperature. 

Today the energy of these gravitons will be comparable, but slightly smaller than the energy of a CMB photon\footnote{The photons that have been until then in thermal equilibrium in the early universe will have  their energy boosted by subsequent annihilation of other species.} Namely, it will be of order $\sim 10^{-4}$ eV with a corresponding  frequency of  $\sim 10$ GHz.

If inflation wasn't too far from the Planck era we might be able to detect the CGB  by measuring its $\Delta N_{\rm eff,g}$, its contribution to the light particles cosmic background. Otherwise, the argument can be reversed to set a new model-independent limit on the epoch of inflation.  An upper limit on $\Delta N_{\rm eff,g}$ (where here we consider just the contribution of these specific gravitons that are formed via annihilation and are today at $\sim 10$GHz implies a lower limit on the energy scale in which inflation took place.

\subsubsection{Annihilation CGB with no inflation}

The situation is much more interesting if inflation didn't take place. In this case the relic gravitons will have a non negligible $\Omega_{\rm g}$. The above estimate, that ignores gravitons annihilating back to regular particles, suggests  that when this is also taken into account the gravitons are in thermal equilibrium with the rest of the universe. In such case we expect 
\begin{equation}
\Omega_{\rm g} \lesssim \frac{\Omega_{\rm CMB}}{g_{\rm pl}} \ , 
\end{equation}
where ${g_{\rm pl}}$ is the unknown number of degrees of freedom (particle species multiplied by the spin factor) at the Planck time. ${g_{\rm pl}}$ is most likely  large and hence $\Omega_{\rm g}$  is  too small to resolve the Hubble tension. 

\bibliographystyle{apsrev4-2}
\bibliography{gravitons}

\begin{thebibliography}{30}%
\makeatletter
\providecommand \@ifxundefined [1]{%
 \@ifx{#1\undefined}
}%
\providecommand \@ifnum [1]{%
 \ifnum #1\expandafter \@firstoftwo
 \else \expandafter \@secondoftwo
 \fi
}%
\providecommand \@ifx [1]{%
 \ifx #1\expandafter \@firstoftwo
 \else \expandafter \@secondoftwo
 \fi
}%
\providecommand \natexlab [1]{#1}%
\providecommand \enquote  [1]{``#1''}%
\providecommand \bibnamefont  [1]{#1}%
\providecommand \bibfnamefont [1]{#1}%
\providecommand \citenamefont [1]{#1}%
\providecommand \href@noop [0]{\@secondoftwo}%
\providecommand \href [0]{\begingroup \@sanitize@url \@href}%
\providecommand \@href[1]{\@@startlink{#1}\@@href}%
\providecommand \@@href[1]{\endgroup#1\@@endlink}%
\providecommand \@sanitize@url [0]{\catcode `\\12\catcode `\$12\catcode
  `\&12\catcode `\#12\catcode `\^12\catcode `\_12\catcode `\%12\relax}%
\providecommand \@@startlink[1]{}%
\providecommand \@@endlink[0]{}%
\providecommand \url  [0]{\begingroup\@sanitize@url \@url }%
\providecommand \@url [1]{\endgroup\@href {#1}{\urlprefix }}%
\providecommand \urlprefix  [0]{URL }%
\providecommand \Eprint [0]{\href }%
\providecommand \doibase [0]{https://doi.org/}%
\providecommand \selectlanguage [0]{\@gobble}%
\providecommand \bibinfo  [0]{\@secondoftwo}%
\providecommand \bibfield  [0]{\@secondoftwo}%
\providecommand \translation [1]{[#1]}%
\providecommand \BibitemOpen [0]{}%
\providecommand \bibitemStop [0]{}%
\providecommand \bibitemNoStop [0]{.\EOS\space}%
\providecommand \EOS [0]{\spacefactor3000\relax}%
\providecommand \BibitemShut  [1]{\csname bibitem#1\endcsname}%
\let\auto@bib@innerbib\@empty
\bibitem [{\citenamefont {{Hawking}}(1974)}]{Hawking1974}%
  \BibitemOpen
  \bibfield  {author} {\bibinfo {author} {\bibfnamefont {S.~W.}\ \bibnamefont
  {{Hawking}}},\ }\href {https://doi.org/10.1038/248030a0} {\bibfield
  {journal} {\bibinfo  {journal} {\nat}\ }\textbf {\bibinfo {volume} {248}},\
  \bibinfo {pages} {30} (\bibinfo {year} {1974})}\BibitemShut {NoStop}%
\bibitem [{\citenamefont {{Hawking}}(1975)}]{Hawking}%
  \BibitemOpen
  \bibfield  {author} {\bibinfo {author} {\bibfnamefont {S.~W.}\ \bibnamefont
  {{Hawking}}},\ }\href {https://doi.org/10.1007/BF02345020} {\bibfield
  {journal} {\bibinfo  {journal} {Communications in Mathematical Physics}\
  }\textbf {\bibinfo {volume} {43}},\ \bibinfo {pages} {199} (\bibinfo {year}
  {1975})}\BibitemShut {NoStop}%
\bibitem [{\citenamefont {{Weinfurtner}}\ \emph {et~al.}(2011)\citenamefont
  {{Weinfurtner}}, \citenamefont {{Tedford}}, \citenamefont {{Penrice}},
  \citenamefont {{Unruh}},\ and\ \citenamefont {{Lawrence}}}]{Weinfurtner2011}%
  \BibitemOpen
  \bibfield  {author} {\bibinfo {author} {\bibfnamefont {S.}~\bibnamefont
  {{Weinfurtner}}}, \bibinfo {author} {\bibfnamefont {E.~W.}\ \bibnamefont
  {{Tedford}}}, \bibinfo {author} {\bibfnamefont {M.~C.~J.}\ \bibnamefont
  {{Penrice}}}, \bibinfo {author} {\bibfnamefont {W.~G.}\ \bibnamefont
  {{Unruh}}},\ and\ \bibinfo {author} {\bibfnamefont {G.~A.}\ \bibnamefont
  {{Lawrence}}},\ }\href {https://doi.org/10.1103/PhysRevLett.106.021302}
  {\bibfield  {journal} {\bibinfo  {journal} {\prl}\ }\textbf {\bibinfo
  {volume} {106}},\ \bibinfo {eid} {021302} (\bibinfo {year} {2011})},\ \Eprint
  {https://arxiv.org/abs/1008.1911} {arXiv:1008.1911 [gr-qc]} \BibitemShut
  {NoStop}%
\bibitem [{\citenamefont {Drori}\ \emph {et~al.}(2019)\citenamefont {Drori},
  \citenamefont {Rosenberg}, \citenamefont {Bermudez}, \citenamefont
  {Silberberg},\ and\ \citenamefont {Leonhardt}}]{analogue}%
  \BibitemOpen
  \bibfield  {author} {\bibinfo {author} {\bibfnamefont {J.}~\bibnamefont
  {Drori}}, \bibinfo {author} {\bibfnamefont {Y.}~\bibnamefont {Rosenberg}},
  \bibinfo {author} {\bibfnamefont {D.}~\bibnamefont {Bermudez}}, \bibinfo
  {author} {\bibfnamefont {Y.}~\bibnamefont {Silberberg}},\ and\ \bibinfo
  {author} {\bibfnamefont {U.}~\bibnamefont {Leonhardt}},\ }\href
  {https://doi.org/10.1103/PhysRevLett.122.010404} {\bibfield  {journal}
  {\bibinfo  {journal} {Phys. Rev. Lett.}\ }\textbf {\bibinfo {volume} {122}},\
  \bibinfo {pages} {010404} (\bibinfo {year} {2019})}\BibitemShut {NoStop}%
\bibitem [{\citenamefont {{Hooper}}\ \emph {et~al.}(2019)\citenamefont
  {{Hooper}}, \citenamefont {{Krnjaic}},\ and\ \citenamefont
  {{McDermott}}}]{Hooper}%
  \BibitemOpen
  \bibfield  {author} {\bibinfo {author} {\bibfnamefont {D.}~\bibnamefont
  {{Hooper}}}, \bibinfo {author} {\bibfnamefont {G.}~\bibnamefont
  {{Krnjaic}}},\ and\ \bibinfo {author} {\bibfnamefont {S.~D.}\ \bibnamefont
  {{McDermott}}},\ }\href {https://doi.org/10.1007/JHEP08(2019)001} {\bibfield
  {journal} {\bibinfo  {journal} {Journal of High Energy Physics}\ }\textbf
  {\bibinfo {volume} {2019}},\ \bibinfo {eid} {1} (\bibinfo {year} {2019})},\
  \Eprint {https://arxiv.org/abs/1905.01301} {arXiv:1905.01301 [hep-ph]}
  \BibitemShut {NoStop}%
\bibitem [{\citenamefont {{Masina}}(2020)}]{Masina}%
  \BibitemOpen
  \bibfield  {author} {\bibinfo {author} {\bibfnamefont {I.}~\bibnamefont
  {{Masina}}},\ }\href {https://doi.org/10.1140/epjp/s13360-020-00564-9}
  {\bibfield  {journal} {\bibinfo  {journal} {European Physical Journal Plus}\
  }\textbf {\bibinfo {volume} {135}},\ \bibinfo {eid} {552} (\bibinfo {year}
  {2020})},\ \Eprint {https://arxiv.org/abs/2004.04740} {arXiv:2004.04740
  [hep-ph]} \BibitemShut {NoStop}%
\bibitem [{\citenamefont {{Arbey}}\ \emph {et~al.}(2021)\citenamefont
  {{Arbey}}, \citenamefont {{Auffinger}}, \citenamefont {{Sandick}},
  \citenamefont {{Shams Es Haghi}},\ and\ \citenamefont {{Sinha}}}]{Arbey+}%
  \BibitemOpen
  \bibfield  {author} {\bibinfo {author} {\bibfnamefont {A.}~\bibnamefont
  {{Arbey}}}, \bibinfo {author} {\bibfnamefont {J.}~\bibnamefont
  {{Auffinger}}}, \bibinfo {author} {\bibfnamefont {P.}~\bibnamefont
  {{Sandick}}}, \bibinfo {author} {\bibfnamefont {B.}~\bibnamefont {{Shams Es
  Haghi}}},\ and\ \bibinfo {author} {\bibfnamefont {K.}~\bibnamefont
  {{Sinha}}},\ }\href {https://doi.org/10.1103/PhysRevD.103.123549} {\bibfield
  {journal} {\bibinfo  {journal} {\prd}\ }\textbf {\bibinfo {volume} {103}},\
  \bibinfo {eid} {123549} (\bibinfo {year} {2021})},\ \Eprint
  {https://arxiv.org/abs/2104.04051} {arXiv:2104.04051 [astro-ph.CO]}
  \BibitemShut {NoStop}%
\bibitem [{\citenamefont {{Auffinger}}(2022)}]{Auffinger}%
  \BibitemOpen
  \bibfield  {author} {\bibinfo {author} {\bibfnamefont {J.}~\bibnamefont
  {{Auffinger}}},\ }\href@noop {} {\bibfield  {journal} {\bibinfo  {journal}
  {arXiv e-prints}\ ,\ \bibinfo {eid} {arXiv:2206.02672}} (\bibinfo {year}
  {2022})},\ \Eprint {https://arxiv.org/abs/2206.02672} {arXiv:2206.02672
  [astro-ph.CO]} \BibitemShut {NoStop}%
\bibitem [{\citenamefont {{Verde}}\ \emph {et~al.}(2019)\citenamefont
  {{Verde}}, \citenamefont {{Treu}},\ and\ \citenamefont
  {{Riess}}}]{VerdeTreu}%
  \BibitemOpen
  \bibfield  {author} {\bibinfo {author} {\bibfnamefont {L.}~\bibnamefont
  {{Verde}}}, \bibinfo {author} {\bibfnamefont {T.}~\bibnamefont {{Treu}}},\
  and\ \bibinfo {author} {\bibfnamefont {A.~G.}\ \bibnamefont {{Riess}}},\
  }\href {https://doi.org/10.1038/s41550-019-0902-0} {\bibfield  {journal}
  {\bibinfo  {journal} {Nature Astronomy}\ }\textbf {\bibinfo {volume} {3}},\
  \bibinfo {pages} {891} (\bibinfo {year} {2019})},\ \Eprint
  {https://arxiv.org/abs/1907.10625} {arXiv:1907.10625 [astro-ph.CO]}
  \BibitemShut {NoStop}%
\bibitem [{\citenamefont {Aghanim}\ \emph {et~al.}(2020)\citenamefont
  {Aghanim}, \citenamefont {Akrami}, \citenamefont {Ashdown}, \citenamefont
  {Aumont}, \citenamefont {Baccigalupi}, \citenamefont {Ballardini},
  \citenamefont {Banday}, \citenamefont {Barreiro}, \citenamefont {Bartolo},\
  and\ \citenamefont {et~al.}}]{Planck18}%
  \BibitemOpen
  \bibfield  {author} {\bibinfo {author} {\bibfnamefont {N.}~\bibnamefont
  {Aghanim}}, \bibinfo {author} {\bibfnamefont {Y.}~\bibnamefont {Akrami}},
  \bibinfo {author} {\bibfnamefont {M.}~\bibnamefont {Ashdown}}, \bibinfo
  {author} {\bibfnamefont {J.}~\bibnamefont {Aumont}}, \bibinfo {author}
  {\bibfnamefont {C.}~\bibnamefont {Baccigalupi}}, \bibinfo {author}
  {\bibfnamefont {M.}~\bibnamefont {Ballardini}}, \bibinfo {author}
  {\bibfnamefont {A.~J.}\ \bibnamefont {Banday}}, \bibinfo {author}
  {\bibfnamefont {R.~B.}\ \bibnamefont {Barreiro}}, \bibinfo {author}
  {\bibfnamefont {N.}~\bibnamefont {Bartolo}},\ and\ \bibinfo {author}
  {\bibnamefont {et~al.}},\ }\href
  {https://doi.org/10.1051/0004-6361/201833910} {\bibfield  {journal} {\bibinfo
   {journal} {Astronomy \& Astrophysics}\ }\textbf {\bibinfo {volume} {641}},\
  \bibinfo {pages} {A6} (\bibinfo {year} {2020})}\BibitemShut {NoStop}%
\bibitem [{\citenamefont {{Sch{\"o}neberg}}\ \emph {et~al.}(2019)\citenamefont
  {{Sch{\"o}neberg}}, \citenamefont {{Lesgourgues}},\ and\ \citenamefont
  {{Hooper}}}]{NilsBBN}%
  \BibitemOpen
  \bibfield  {author} {\bibinfo {author} {\bibfnamefont {N.}~\bibnamefont
  {{Sch{\"o}neberg}}}, \bibinfo {author} {\bibfnamefont {J.}~\bibnamefont
  {{Lesgourgues}}},\ and\ \bibinfo {author} {\bibfnamefont {D.~C.}\
  \bibnamefont {{Hooper}}},\ }\href
  {https://doi.org/10.1088/1475-7516/2019/10/029} {\bibfield  {journal}
  {\bibinfo  {journal} {JCAP}\ }\textbf {\bibinfo {volume} {2019}}\bibfield
  {number} {\bibinfo  {number} { (10)},\ \bibinfo {eid} {029}},\ }\Eprint
  {https://arxiv.org/abs/1907.11594} {arXiv:1907.11594 [astro-ph.CO]}
  \BibitemShut {NoStop}%
\bibitem [{\citenamefont {{Sch{\"o}neberg}}\ \emph {et~al.}(2022)\citenamefont
  {{Sch{\"o}neberg}}, \citenamefont {{Verde}}, \citenamefont
  {{Gil-Mar{\'\i}n}},\ and\ \citenamefont {{Brieden}}}]{BBNNils}%
  \BibitemOpen
  \bibfield  {author} {\bibinfo {author} {\bibfnamefont {N.}~\bibnamefont
  {{Sch{\"o}neberg}}}, \bibinfo {author} {\bibfnamefont {L.}~\bibnamefont
  {{Verde}}}, \bibinfo {author} {\bibfnamefont {H.}~\bibnamefont
  {{Gil-Mar{\'\i}n}}},\ and\ \bibinfo {author} {\bibfnamefont {S.}~\bibnamefont
  {{Brieden}}},\ }\href@noop {} {\bibfield  {journal} {\bibinfo  {journal}
  {arXiv e-prints}\ ,\ \bibinfo {eid} {arXiv:2209.14330}} (\bibinfo {year}
  {2022})},\ \Eprint {https://arxiv.org/abs/2209.14330} {arXiv:2209.14330
  [astro-ph.CO]} \BibitemShut {NoStop}%
\bibitem [{\citenamefont {{Gomez}}\ and\ \citenamefont
  {{Jimenez}}(2022)}]{Fisher}%
  \BibitemOpen
  \bibfield  {author} {\bibinfo {author} {\bibfnamefont {C.}~\bibnamefont
  {{Gomez}}}\ and\ \bibinfo {author} {\bibfnamefont {R.}~\bibnamefont
  {{Jimenez}}},\ }\href {https://doi.org/10.1016/j.dark.2022.101035} {\bibfield
   {journal} {\bibinfo  {journal} {Physics of the Dark Universe}\ }\textbf
  {\bibinfo {volume} {36}},\ \bibinfo {eid} {101035} (\bibinfo {year}
  {2022})},\ \Eprint {https://arxiv.org/abs/2111.05380} {arXiv:2111.05380
  [hep-th]} \BibitemShut {NoStop}%
\bibitem [{\citenamefont {{Page}}(1976{\natexlab{a}})}]{Page1976}%
  \BibitemOpen
  \bibfield  {author} {\bibinfo {author} {\bibfnamefont {D.~N.}\ \bibnamefont
  {{Page}}},\ }\href {https://doi.org/10.1103/PhysRevD.13.198} {\bibfield
  {journal} {\bibinfo  {journal} {\prd}\ }\textbf {\bibinfo {volume} {13}},\
  \bibinfo {pages} {198} (\bibinfo {year} {1976}{\natexlab{a}})}\BibitemShut
  {NoStop}%
\bibitem [{\citenamefont {{Borsanyi}}\ \emph {et~al.}(2016)\citenamefont
  {{Borsanyi}}, \citenamefont {{Fodor}}, \citenamefont {{Kampert}},
  \citenamefont {{Katz}}, \citenamefont {{Kawanai}}, \citenamefont {{Kovacs}},
  \citenamefont {{Mages}}, \citenamefont {{Pasztor}}, \citenamefont
  {{Pittler}}, \citenamefont {{Redondo}}, \citenamefont {{Ringwald}},\ and\
  \citenamefont {{Szabo}}}]{Borsanyi2016}%
  \BibitemOpen
  \bibfield  {author} {\bibinfo {author} {\bibfnamefont {S.}~\bibnamefont
  {{Borsanyi}}}, \bibinfo {author} {\bibfnamefont {Z.}~\bibnamefont {{Fodor}}},
  \bibinfo {author} {\bibfnamefont {K.~H.}\ \bibnamefont {{Kampert}}}, \bibinfo
  {author} {\bibfnamefont {S.~D.}\ \bibnamefont {{Katz}}}, \bibinfo {author}
  {\bibfnamefont {T.}~\bibnamefont {{Kawanai}}}, \bibinfo {author}
  {\bibfnamefont {T.~G.}\ \bibnamefont {{Kovacs}}}, \bibinfo {author}
  {\bibfnamefont {S.~W.}\ \bibnamefont {{Mages}}}, \bibinfo {author}
  {\bibfnamefont {A.}~\bibnamefont {{Pasztor}}}, \bibinfo {author}
  {\bibfnamefont {F.}~\bibnamefont {{Pittler}}}, \bibinfo {author}
  {\bibfnamefont {J.}~\bibnamefont {{Redondo}}}, \bibinfo {author}
  {\bibfnamefont {A.}~\bibnamefont {{Ringwald}}},\ and\ \bibinfo {author}
  {\bibfnamefont {K.~K.}\ \bibnamefont {{Szabo}}},\ }\href@noop {} {\bibfield
  {journal} {\bibinfo  {journal} {arXiv e-prints}\ ,\ \bibinfo {eid}
  {arXiv:1606.07494}} (\bibinfo {year} {2016})},\ \Eprint
  {https://arxiv.org/abs/1606.07494} {arXiv:1606.07494 [hep-lat]} \BibitemShut
  {NoStop}%
\bibitem [{\citenamefont {{Page}}(1976{\natexlab{b}})}]{Page1976R}%
  \BibitemOpen
  \bibfield  {author} {\bibinfo {author} {\bibfnamefont {D.~N.}\ \bibnamefont
  {{Page}}},\ }\href {https://doi.org/10.1103/PhysRevD.14.3260} {\bibfield
  {journal} {\bibinfo  {journal} {\prd}\ }\textbf {\bibinfo {volume} {14}},\
  \bibinfo {pages} {3260} (\bibinfo {year} {1976}{\natexlab{b}})}\BibitemShut
  {NoStop}%
\bibitem [{\citenamefont {Weinberg}(1965)}]{Weinberg1965}%
  \BibitemOpen
  \bibfield  {author} {\bibinfo {author} {\bibfnamefont {S.}~\bibnamefont
  {Weinberg}},\ }\href {https://doi.org/10.1103/PhysRev.140.B516} {\bibfield
  {journal} {\bibinfo  {journal} {Phys. Rev.}\ }\textbf {\bibinfo {volume}
  {140}},\ \bibinfo {pages} {B516} (\bibinfo {year} {1965})}\BibitemShut
  {NoStop}%
\bibitem [{\citenamefont {{Morrison}}\ \emph {et~al.}(2019)\citenamefont
  {{Morrison}}, \citenamefont {{Profumo}},\ and\ \citenamefont
  {{Yu}}}]{Profumo}%
  \BibitemOpen
  \bibfield  {author} {\bibinfo {author} {\bibfnamefont {L.}~\bibnamefont
  {{Morrison}}}, \bibinfo {author} {\bibfnamefont {S.}~\bibnamefont
  {{Profumo}}},\ and\ \bibinfo {author} {\bibfnamefont {Y.}~\bibnamefont
  {{Yu}}},\ }\href {https://doi.org/10.1088/1475-7516/2019/05/005} {\bibfield
  {journal} {\bibinfo  {journal} {JCAP}\ }\textbf {\bibinfo {volume}
  {2019}}\bibfield  {number} {\bibinfo  {number} { (5)},\ \bibinfo {eid}
  {005}},\ }\Eprint {https://arxiv.org/abs/1812.10606} {arXiv:1812.10606
  [astro-ph.CO]} \BibitemShut {NoStop}%
\bibitem [{\citenamefont {{Sch{\"o}neberg}}\ \emph {et~al.}(2021)\citenamefont
  {{Sch{\"o}neberg}}, \citenamefont {{Abell{\'a}n}}, \citenamefont {{P{\'e}rez
  S{\'a}nchez}}, \citenamefont {{Witte}}, \citenamefont {{Poulin}},\ and\
  \citenamefont {{Lesgourgues}}}]{olympics}%
  \BibitemOpen
  \bibfield  {author} {\bibinfo {author} {\bibfnamefont {N.}~\bibnamefont
  {{Sch{\"o}neberg}}}, \bibinfo {author} {\bibfnamefont {G.~F.}\ \bibnamefont
  {{Abell{\'a}n}}}, \bibinfo {author} {\bibfnamefont {A.}~\bibnamefont
  {{P{\'e}rez S{\'a}nchez}}}, \bibinfo {author} {\bibfnamefont {S.~J.}\
  \bibnamefont {{Witte}}}, \bibinfo {author} {\bibfnamefont {V.}~\bibnamefont
  {{Poulin}}},\ and\ \bibinfo {author} {\bibfnamefont {J.}~\bibnamefont
  {{Lesgourgues}}},\ }\href@noop {} {\bibfield  {journal} {\bibinfo  {journal}
  {arXiv e-prints}\ ,\ \bibinfo {eid} {arXiv:2107.10291}} (\bibinfo {year}
  {2021})},\ \Eprint {https://arxiv.org/abs/2107.10291} {arXiv:2107.10291
  [astro-ph.CO]} \BibitemShut {NoStop}%
\bibitem [{\citenamefont {{Heavens}}\ \emph {et~al.}(2014)\citenamefont
  {{Heavens}}, \citenamefont {{Jimenez}},\ and\ \citenamefont
  {{Verde}}}]{ruler}%
  \BibitemOpen
  \bibfield  {author} {\bibinfo {author} {\bibfnamefont {A.}~\bibnamefont
  {{Heavens}}}, \bibinfo {author} {\bibfnamefont {R.}~\bibnamefont
  {{Jimenez}}},\ and\ \bibinfo {author} {\bibfnamefont {L.}~\bibnamefont
  {{Verde}}},\ }\href {https://doi.org/10.1103/PhysRevLett.113.241302}
  {\bibfield  {journal} {\bibinfo  {journal} {PRL}\ }\textbf {\bibinfo {volume}
  {113}},\ \bibinfo {eid} {241302} (\bibinfo {year} {2014})},\ \Eprint
  {https://arxiv.org/abs/1409.6217} {arXiv:1409.6217 [astro-ph.CO]}
  \BibitemShut {NoStop}%
\bibitem [{\citenamefont {{Verde}}\ \emph {et~al.}(2017)\citenamefont
  {{Verde}}, \citenamefont {{Bernal}}, \citenamefont {{Heavens}},\ and\
  \citenamefont {{Jimenez}}}]{bernal}%
  \BibitemOpen
  \bibfield  {author} {\bibinfo {author} {\bibfnamefont {L.}~\bibnamefont
  {{Verde}}}, \bibinfo {author} {\bibfnamefont {J.~L.}\ \bibnamefont
  {{Bernal}}}, \bibinfo {author} {\bibfnamefont {A.~F.}\ \bibnamefont
  {{Heavens}}},\ and\ \bibinfo {author} {\bibfnamefont {R.}~\bibnamefont
  {{Jimenez}}},\ }\href {https://doi.org/10.1093/mnras/stx116} {\bibfield
  {journal} {\bibinfo  {journal} {MNRAS}\ }\textbf {\bibinfo {volume} {467}},\
  \bibinfo {pages} {731} (\bibinfo {year} {2017})},\ \Eprint
  {https://arxiv.org/abs/1607.05297} {arXiv:1607.05297 [astro-ph.CO]}
  \BibitemShut {NoStop}%
\bibitem [{\citenamefont {{Kolb}}\ and\ \citenamefont
  {{Turner}}(1990)}]{KolbTurner}%
  \BibitemOpen
  \bibfield  {author} {\bibinfo {author} {\bibfnamefont {E.~W.}\ \bibnamefont
  {{Kolb}}}\ and\ \bibinfo {author} {\bibfnamefont {M.~S.}\ \bibnamefont
  {{Turner}}},\ }\href@noop {} {\emph {\bibinfo {title} {{The early
  universe}}}},\ Vol.~\bibinfo {volume} {69}\ (\bibinfo  {publisher} {Univ.
  Chicago Press, Chicago},\ \bibinfo {year} {1990})\BibitemShut {NoStop}%
\bibitem [{\citenamefont {{Hooper}}\ \emph {et~al.}(2020)\citenamefont
  {{Hooper}}, \citenamefont {{Krnjaic}}, \citenamefont {{March-Russell}},
  \citenamefont {{McDermott}},\ and\ \citenamefont
  {{Petrossian-Byrne}}}]{HooperKerr}%
  \BibitemOpen
  \bibfield  {author} {\bibinfo {author} {\bibfnamefont {D.}~\bibnamefont
  {{Hooper}}}, \bibinfo {author} {\bibfnamefont {G.}~\bibnamefont {{Krnjaic}}},
  \bibinfo {author} {\bibfnamefont {J.}~\bibnamefont {{March-Russell}}},
  \bibinfo {author} {\bibfnamefont {S.~D.}\ \bibnamefont {{McDermott}}},\ and\
  \bibinfo {author} {\bibfnamefont {R.}~\bibnamefont {{Petrossian-Byrne}}},\
  }\href@noop {} {\bibfield  {journal} {\bibinfo  {journal} {arXiv e-prints}\
  ,\ \bibinfo {eid} {arXiv:2004.00618}} (\bibinfo {year} {2020})},\ \Eprint
  {https://arxiv.org/abs/2004.00618} {arXiv:2004.00618 [astro-ph.CO]}
  \BibitemShut {NoStop}%
\bibitem [{\citenamefont {{Berlin}}\ \emph {et~al.}(2022)\citenamefont
  {{Berlin}}, \citenamefont {{Blas}}, \citenamefont {{D'Agnolo}}, \citenamefont
  {{Ellis}}, \citenamefont {{Harnik}}, \citenamefont {{Kahn}},\ and\
  \citenamefont {{Sch{\"u}tte-Engel}}}]{Blas}%
  \BibitemOpen
  \bibfield  {author} {\bibinfo {author} {\bibfnamefont {A.}~\bibnamefont
  {{Berlin}}}, \bibinfo {author} {\bibfnamefont {D.}~\bibnamefont {{Blas}}},
  \bibinfo {author} {\bibfnamefont {R.~T.}\ \bibnamefont {{D'Agnolo}}},
  \bibinfo {author} {\bibfnamefont {S.~A.~R.}\ \bibnamefont {{Ellis}}},
  \bibinfo {author} {\bibfnamefont {R.}~\bibnamefont {{Harnik}}}, \bibinfo
  {author} {\bibfnamefont {Y.}~\bibnamefont {{Kahn}}},\ and\ \bibinfo {author}
  {\bibfnamefont {J.}~\bibnamefont {{Sch{\"u}tte-Engel}}},\ }\href
  {https://doi.org/10.1103/PhysRevD.105.116011} {\bibfield  {journal} {\bibinfo
   {journal} {PRD}\ }\textbf {\bibinfo {volume} {105}},\ \bibinfo {eid}
  {116011} (\bibinfo {year} {2022})},\ \Eprint
  {https://arxiv.org/abs/2112.11465} {arXiv:2112.11465 [hep-ph]} \BibitemShut
  {NoStop}%
\bibitem [{\citenamefont {Brieden}\ \emph {et~al.}(2022)\citenamefont
  {Brieden}, \citenamefont {Gil-Mar\'\i{}n},\ and\ \citenamefont
  {Verde}}]{shapefit}%
  \BibitemOpen
  \bibfield  {author} {\bibinfo {author} {\bibfnamefont {S.}~\bibnamefont
  {Brieden}}, \bibinfo {author} {\bibfnamefont {H.}~\bibnamefont
  {Gil-Mar\'\i{}n}},\ and\ \bibinfo {author} {\bibfnamefont {L.}~\bibnamefont
  {Verde}},\ }\href {https://doi.org/10.1088/1475-7516/2022/08/024} {\bibfield
  {journal} {\bibinfo  {journal} {JCAP}\ }\textbf {\bibinfo {volume}
  {08}}\bibfield  {number} {\bibinfo  {number} { (08)},\ \bibinfo {pages}
  {024}},\ }\Eprint {https://arxiv.org/abs/2204.11868} {arXiv:2204.11868
  [astro-ph.CO]} \BibitemShut {NoStop}%
\bibitem [{\citenamefont {{Vladimirov}}(1963)}]{Vladimirov1963}%
  \BibitemOpen
  \bibfield  {author} {\bibinfo {author} {\bibfnamefont {Y.~S.}\ \bibnamefont
  {{Vladimirov}}},\ }\href@noop {} {\bibfield  {journal} {\bibinfo  {journal}
  {Soviet Journal of Experimental and Theoretical Physics}\ }\textbf {\bibinfo
  {volume} {16}},\ \bibinfo {pages} {65} (\bibinfo {year} {1963})}\BibitemShut
  {NoStop}%
\bibitem [{\citenamefont {{Papini}}\ and\ \citenamefont
  {{Valluri}}(1977)}]{Papini1977}%
  \BibitemOpen
  \bibfield  {author} {\bibinfo {author} {\bibfnamefont {G.}~\bibnamefont
  {{Papini}}}\ and\ \bibinfo {author} {\bibfnamefont {S.~R.}\ \bibnamefont
  {{Valluri}}},\ }\href {https://doi.org/10.1016/0370-1573(77)90006-0}
  {\bibfield  {journal} {\bibinfo  {journal} {Physics Reports}\ }\textbf
  {\bibinfo {volume} {33C}},\ \bibinfo {pages} {51} (\bibinfo {year}
  {1977})}\BibitemShut {NoStop}%
\bibitem [{\citenamefont {{Zel'dovich}}(1967)}]{Zel'dovich1967}%
  \BibitemOpen
  \bibfield  {author} {\bibinfo {author} {\bibfnamefont {Y.~B.}\ \bibnamefont
  {{Zel'dovich}}},\ }\href {https://doi.org/10.1070/PU1967v009n04ABEH003014}
  {\bibfield  {journal} {\bibinfo  {journal} {Soviet Physics Uspekhi}\ }\textbf
  {\bibinfo {volume} {9}},\ \bibinfo {pages} {602} (\bibinfo {year}
  {1967})}\BibitemShut {NoStop}%
\bibitem [{\citenamefont {{Smolin}}(1985)}]{Smolin1985}%
  \BibitemOpen
  \bibfield  {author} {\bibinfo {author} {\bibfnamefont {L.}~\bibnamefont
  {{Smolin}}},\ }\href {https://doi.org/10.1007/BF00761902} {\bibfield
  {journal} {\bibinfo  {journal} {General Relativity and Gravitation}\ }\textbf
  {\bibinfo {volume} {3}},\ \bibinfo {pages} {17} (\bibinfo {year}
  {1985})}\BibitemShut {NoStop}%
\bibitem [{\citenamefont {{Matzner}}(1968)}]{Matzner}%
  \BibitemOpen
  \bibfield  {author} {\bibinfo {author} {\bibfnamefont {R.~A.}\ \bibnamefont
  {{Matzner}}},\ }\href {https://doi.org/10.1086/149831} {\bibfield  {journal}
  {\bibinfo  {journal} {\apj}\ }\textbf {\bibinfo {volume} {154}},\ \bibinfo
  {pages} {1123} (\bibinfo {year} {1968})}\BibitemShut {NoStop}%
\end{thebibliography}

\providecommand{\noopsort}[1]{}\providecommand{\singleletter}[1]{#1}%

\end{document}